\documentclass[aps,prd,twocolumn,groupedaddress,showpacs,amsmath,amssymb]{revtex4-1}

\usepackage{graphicx}
\usepackage{bm,color}

\newcommand{\be}{\begin{eqnarray}}
\newcommand{\ee}{\end{eqnarray}}

\def\tchi{\tilde{\chi}}
\def\tPhi{\tilde{\Phi}}
\def\vr{\vec{r}}

\begin{document}

\title{Extended stream functions for dynamics of Bose-Einstein condensations: \\
Snake instability of dark soliton in ultra-cold atoms as an example
}
\author{Takahiro Orito}
\author{Ikuo Ichinose}
\affiliation{Department of Applied Physics, Nagoya Institute of Technology, Nagoya, 466-8555, Japan}

\date{\today}

\begin{abstract}
In this paper, we formulate extended stream functions (ESFs) to describe the dynamics of 
Bose-Einstein condensations in the two-dimensional space.
The ordinary stream function is applicable only for stationary and incompressible
superfluids, whereas the ESFs can describe the dynamics of compressible and 
non-stationary superfluids.
The ESFs are composed of two stream functions, i.e.,
one describes the compressible density modulations and the other the incompressible rotational
superflow.
As an application, we study the snake instability of the dark soliton in a rectangular
potential in detail by the ESFs.
\end{abstract}

\maketitle


\section{Introduction}{\label{intro}}

In the past decade, ultracold atomic systems are one of the most actively
studied fields in physics.
It is expected that these systems can be a quantum simulator for 
various models in quantum physics~\cite{qsimu}.
Another aim to study ultracold atoms is to search for new quantum
phenomena whose existence is theoretically predicted.
These example are the supersolid that has both a diagonal
solid order and an off-diagonal superfluid order~\cite{SS}, bosonic analog of 
fractional quantum Hall effect~\cite{FQHE}, etc.
Because of their versatility and controllability, atomic gas systems are
used for study on dynamics of Bose-Einstein condensates (BEC), in particular,
formation and destruction of topological excitations such as vortices.
Detailed study on such dynamical process was given in 
Refs.~\cite{vortex1,vortex2,vortex3,vortex4}, and it shed a light on the dynamics
of vortices, which is a long standing problem since the late 19th~\cite{hydro}.

In this paper, we focus on two-dimensional (2D) BEC and superfluidity (SF).
Recently, vortex dynamics in annular and cylindrical BEC was studied
by using stream function of the SF~\cite{stF1,stF2}.
In these systems, incompressible and irrotational stationary states of the SF 
can be analytically studied by the stream function.
In order to extend the stream-function formalism to compressible, irrotational
and non-stationary fluid, a pair of stream functions are needed.
That is,  one describes rotational flow, and the other compressible component of flow.
We call them extended stream functions (ESFs) in this paper.
As we show in the rest of this paper, the ESFs, in our definition, are numerically obtained
without any difficulties and they are quite useful for detailed study on SF dynamics.

The present paper is organized as follows.
In Sec.~II, we introduce the ESFs, and their relation to the Gross-Pitaevskii 
equation (GPE) is explained.
In order to verify the reliability and accuracy of the practical manipulation to obtain 
the incompressible and irrotational components of SF flow from the ESFs, 
we apply our formalism for an annular SF.
By a sudden quench of an synthetic magnetic field, a homogeneous BEC 
becomes unstable, and vortex formation takes place.
We describe this evolution of the BEC by the ESFs and compare the results
with those obtained directly from the solution to the GPE.   
In Sec.~III, we apply the ESFs formalism for certain typical phenomenon 
of the SF instability.
As an example, we study the time evolution of a dark soliton~\cite{DS1,DS2,DS3} located in the center
of a rectangular potential~\cite{BV}.
This dark soliton exhibits a snake instability and the central empty region 
changes its shape, and vortices and antivortices are generated there.
We show that the ESFs clearly describe this process of the destruction
of the dark soliton and formation of vortex-antivortex chain.
Section IV is devoted for conclusion.


\section{Extended stream functions}

As we explained in introduction, we mostly study the BEC in the ultra-cold
bosonic atoms by solving the GPE,
\be
(i-\gamma){\partial \psi \over \partial t}
&=&-{1 \over 2}\nabla^2 \psi +V(r)\psi +g|\psi|^2\psi  \nonumber \\
&&-\bar{\mu} \psi -\omega L_z \psi,
\label{GPeq}
\ee
where $\gamma$ is a phenomenological dissipation parameter, 
$V(r)$ is potential,
$g$ is the coupling constant for the repulsion, and $\bar{\mu}$ is the chemical potential.
We have put the atomic mass $m=1$, and also the Planck constant is set $\hbar=1$.
For example, the unit of time $t=1$ corresponds to $7.94$ ms for $^{87}$Rb confined 
by the 2D harmonic trap with $\omega_\bot=20.1\times 2\pi$ Hz~\cite{vortex4}.
We use this energy scale and the unit of time as well as the normalization, such as
$\int dxdy |\psi|^2=1$, in the rest of present study.
For the system under a rotation, $\omega$ is the angular
velocity of the rotation and $L_z$ is the angular momentum in the $z$-direction.
We are in the co-moving frame of the atoms in Eq.~(\ref{GPeq}).
In the rest of this work, we solve the GPE in Eq.~(\ref{GPeq}) for specific
$V(r)$, initial conditions, etc, and study the dynamics of the BEC in detail.
To this end, we introduce the ESFs.
The ESFs are applicable for various physical systems with flow.
In this paper, we use them for study on SF of ultra-cold atoms.

In the previous works using the ordinary stream function, 
incompressible and stationary BEC is considered~\cite{stF1,stF2}.
For compressible and non-stationary flows of SF, 
the continuity equation is given as follows,
\be
{\partial \rho \over \partial t}+\nabla\cdot (\rho \vec{v})=0,
\label{continue}
\ee
where $\rho\equiv |\psi|^2$ and $\vec{v}\equiv 
{1\over 2i\rho}(\psi^\ast \nabla \psi-\psi \nabla\psi^\ast)=\nabla \theta$ 
with $\psi=|\psi|e^{i\theta}$.
For incompressible and stationary BEC, the continuity equation, Eq.~(\ref{continue}), 
reduces $\nabla\cdot \vec{v}=0$.
Therefore in 2D systems, the velocity $\vec{v}$ can be expressed 
as follows by using a function $\chi(\vec{r})$,
which is called stream function,
\be
v_x=-{\partial \chi \over \partial y}, \;\;
v_y={\partial \chi \over \partial x},
\label{chi}
\ee
where $\vr=(x,y)$.
On the other hand
by the velocity potential $\Phi(\vec{r})$, the velocity $\vec{v}$ is expressed as
\be
v_x={\partial \Phi \over \partial x}, \;\;
v_y={\partial \Phi \over \partial y}.
\label{Phi}
\ee
Then, it is useful to introduce the complex potential such as 
$F(z)\equiv \chi(z) +i\Phi(z)$, where $z$ is the complex coordinate
$z=x+iy$.
From Eqs.~(\ref{chi}) and (\ref{Phi}), it is obvious that $F(z)$ satisfies the 
Cauchy-Riemann condition.
As a result, problems with various boundary conditions can be connected
by a conformal mapping.

In this work, we generalize the above stream function in order to study the compressible 
and non-stationary quantum flows, 
$\vec{j}={1\over 2i}(\psi^\ast \nabla \psi-\psi \nabla\psi^\ast)$.
To this end, we introduce a pair of stream functions, $\tchi(\vr,t)$ and $\tPhi(\vr,t)$, the ESFs.
The flow $\vec{j}$ is composed of the incompressible and irrotational (compressible) parts, i.e.,
\be
\vec{j}=\rho\nabla\theta
=(\rho\nabla\theta)_i+(\rho\nabla\theta)_c\equiv \vec{j}_i+\vec{j}_c,
\label{current}
\ee
where $(\rho\nabla\theta)_i$ and $(\rho\nabla\theta)_c$ stand for the incompressible
and irrotational components of the flow, respectively.
Each of the above components is repressed by the ESFs as follows,
\be
(\rho\nabla\theta)_i=\hat{z}\times \nabla\tchi, \;\;
(\rho\nabla\theta)_c=\nabla \tPhi,
\label{flowF}
\ee
where $\hat{z}$ is the unite vector in the $z$-direction.
The following equalities are easily verified,
\be
\nabla\times \vec{j}=\Delta \tchi, \;\;
\nabla\cdot \vec{j}=\Delta\tPhi.
\label{flowF2}
\ee
The continuity equation (\ref{continue}) is expressed as,
\be
{\partial \rho \over \partial t}+\Delta\tPhi=0.
\label{continue2}
\ee
It should be remarked here that the ESFs $\tchi$ and $\tPhi$ 
include the density $\rho$ in their definition, and therefore they are well defined
for configurations including vortices.
Furthermore for the system confined by a trap potential $V(r)$, the BEC has
boundaries at which the density vanishes.
This means that no boundary conditions are needed to obtain $\tchi$ and 
$\tPhi$ as we see later on.

\begin{figure}[h]
\centering
\begin{center}
\includegraphics[width=7cm]{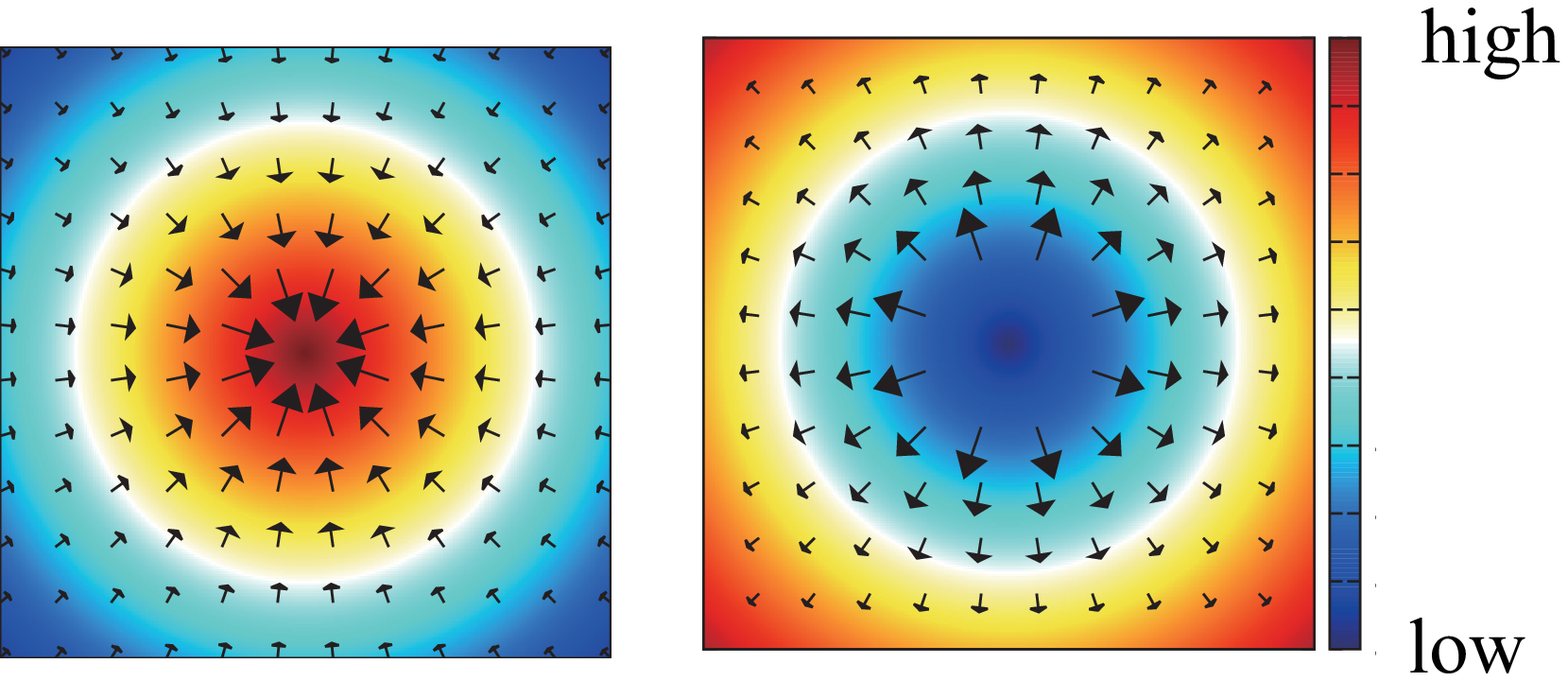} 
\includegraphics[width=7cm]{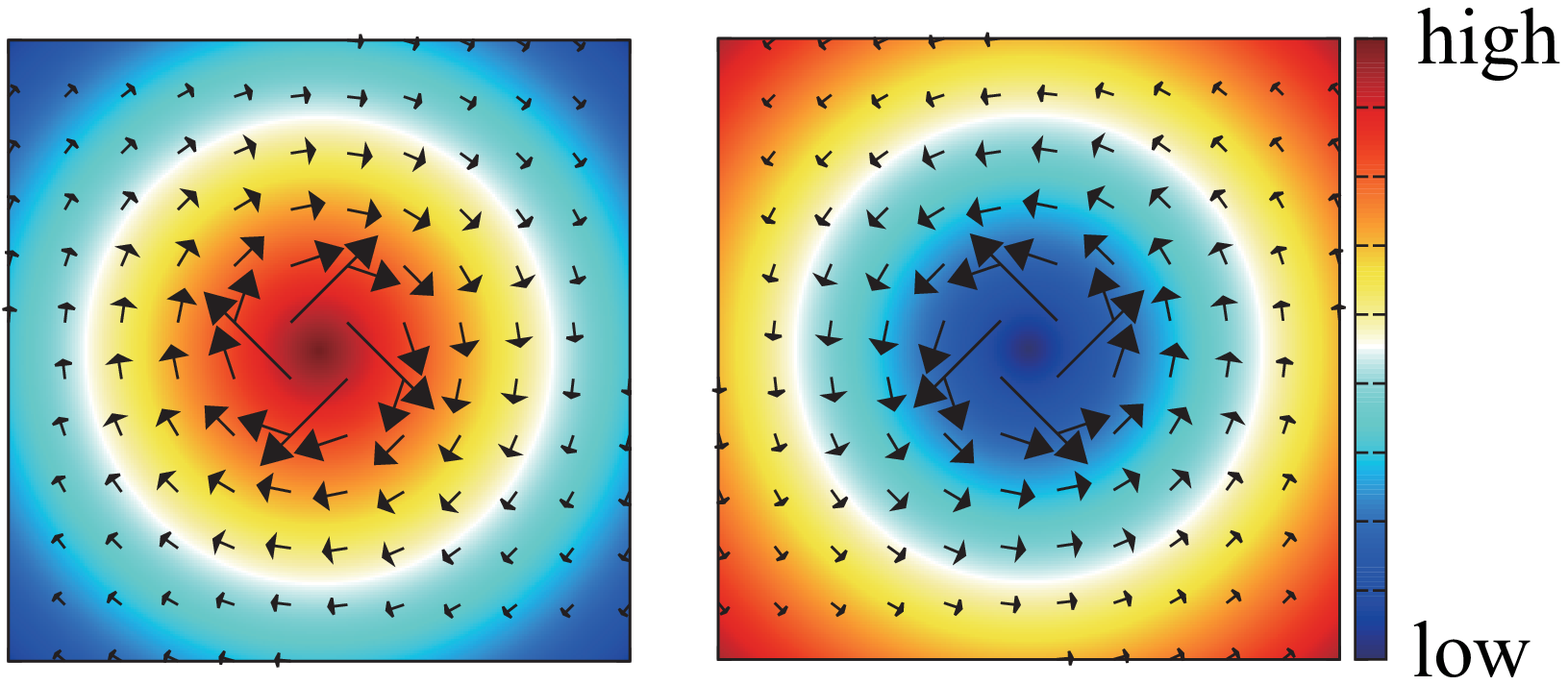}
\end{center}
\caption{Typical profiles of the ESFs, and compressible and 
incompressible flows of the BEC.
$\tPhi$ in upper panels (the compressible component), and
$\tchi$ (the rotational incompressible component) in lower panels.
Arrows indicate flow of each component.
}
\label{phichi}
\end{figure}

From the definition Eq.~(\ref{flowF}), typical profiles of $\tchi$ and $\tPhi$ that
correspond the flows $j_i$ and $j_c$ as shown in Fig.~\ref{phichi}.
These profiles are useful to understand what the numerical calculations of
$\tchi$ and $\tPhi$ indicate in the later discussion.

As we show in the rest of the present paper, the ESFs describe detailed
behavior of the BEC in a transparent way.
In the practical calculation, we obtain $\tchi$ and $\tPhi$ from the current
$\vec{j}(x,y,t)=\vec{j}(\vr,t)$ by using the Fourier/inverse-Fourier transformations,
\be
\tchi&=&-F^{-1}\Big[{1 \over k^2_x+k^2_y}F[\nabla\times \vec{j}]\Big], \nonumber \\
\tPhi&=&-F^{-1}\Big[{1 \over k^2_x+k^2_y}F[\nabla\cdot\vec{j}]\Big],
\label{Fourier}
\ee
where $F$ and $F^{-1}$ denote the operator of the Fourier  and 
inverse-Fourier transformations, respectively.

In order to verify the utility of the ESFs and reliability of the numerical manipulations
of the present formalism, we studied the sudden quench dynamics of BEC confined 
in a 2D torus, which was studied in the previous works.
Abrupt application of an artificial  magnetic field renders homogeneous BEC
unstable.
To see how the BEC evolves in the process, we solve the GPE in 
Eq.~(\ref{GPeq}).
For the numerical calculation, we used the following parameters;
$\gamma=0.03$, $V(r)=30(r-4)^2$, $g=1000$, $\omega=0 \to 1.9$,
and the time and spatial meshes for the calculation are $dx, dy=0.05$ and 
$dt=0.005$, respectively.

\begin{figure}[h]
\centering
\begin{center}
\hspace{-0.5cm}
\includegraphics[width=10cm]{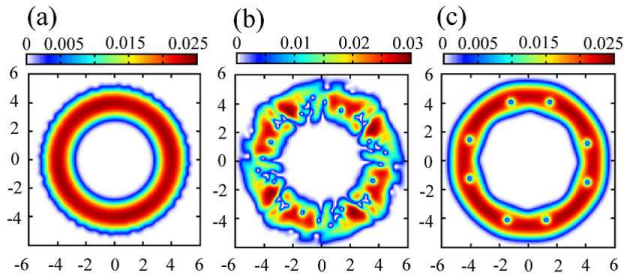}
\end{center}
\vspace{-2cm}
\caption{(a) Evolution of BEC density $\rho$ in a torus after the sudden quench 
of the external magnetic field. ($t=10$.)
(b) Surface ripples propagating the along the surface appear. ($t=14$.)
(c) Stable vortex lattice forms. ($t=100$.)
}
\label{Evolution}
\end{figure}

Evolution of the SF is as follows~\cite{vortex4}.
First, surface ripples propagating along the surface appear and then,
they gradually develop into vortex cores, and finally vortices enter into the BEC and 
a vortex lattice forms.
See Fig.~\ref{Evolution}.
We compared the current $\vec{j}(\vec{r}, t)$ obtained directly from the numerical
solutions of the GPE and that obtained through the ESFs,
$\tchi(\vec{r}, t)$ and $\tPhi(\vec{r}, t)$,  in the final stage (the stage (c) in Fig.~\ref{Evolution}), and 
found that these two calculations are in good agreement.

\begin{figure}[h]
\centering
\begin{center}
\includegraphics[width=8cm]{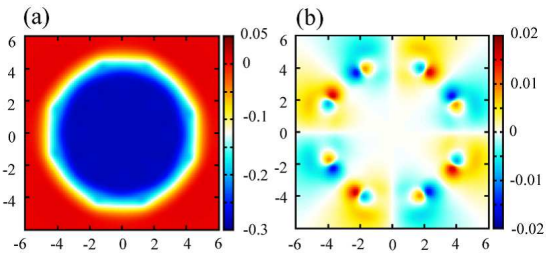}
\end{center}
\vspace{-2cm}
\caption{ESFs (a) $\tchi$ and (b) $\tPhi$ in the final stage of the BEC
evolution after the sudden quench (the stage (c) in Fig.~\ref{Evolution}).
The incompressible component flow described by $\tchi$ circulates around the torus.
On the other hand, the compressible component $\tPhi$ has a pair of source and drain
corresponding to each vortex.
}
\label{streamF}
\end{figure}
\begin{figure}[h]
\centering
\begin{center}
\includegraphics[width=8cm]{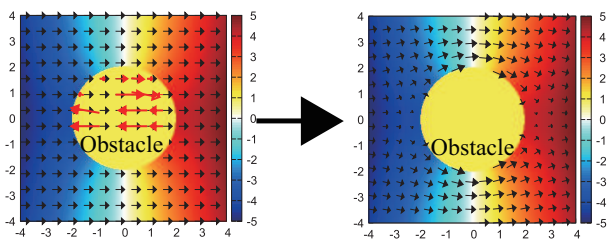}
\end{center}
\caption{ESFs in the vicinity of an obstacle such as a vortex.
The incompressible component described by $\tchi$ is not influenced by the
obstacle, whereas the compressible component $\tPhi$ generates counter-flow
to cancel the $\tchi$-flow inside of the obstacle.
}
\label{obstacle}
\end{figure}

It is quite instructive to see how the ESFs describe the process of vortex lattice
formation in the above quench dynamics.
In Fig.~\ref{streamF}, we first show the ESFs $\tchi(\vec{r})$ and $\tPhi(\vec{r})$ for the 
vortex lattice BEC that formed at the final stage of the evolution. 
Profile of $\tchi(\vec{r})$ shows that $\tchi(\vec{r})$ is a smooth function and 
a steady counter-clockwise SF flow exists inside of the torus.
Behavior of $\tchi(\vec{r})$ is {\em not influenced substantially} by the existence of the vortices
as it describes the incompressible (rotational) component of the SF flow.
On the other hand, $\tPhi(\vec{r})$ exhibits an interesting profile.
High and low-$\tPhi$ regions always appear in a pair, and we call this configuration dimer. 
Around the locations of vortex, a pair of dimer forms, and in the regions between
dimers $\tPhi\sim 0$.
Schematic picture of this $\tPhi$ configuration in the vicinity of an obstacle
(e.g., vortex) is shown in Fig.~\ref{obstacle}.
It shows that 
the compressible component $\vec{j}_c$ cancels the flow of the incompressible
component $\vec{j}_i$ inside of the obstacle.
As a result, the SF density inside of the obstacle is kept vanishing as
$\rho \nabla \theta=\vec{j}=\vec{j}_c+\vec{j}_i\simeq 0$.

\begin{figure}[h]
\centering
\begin{center}
\includegraphics[width=8.7cm]{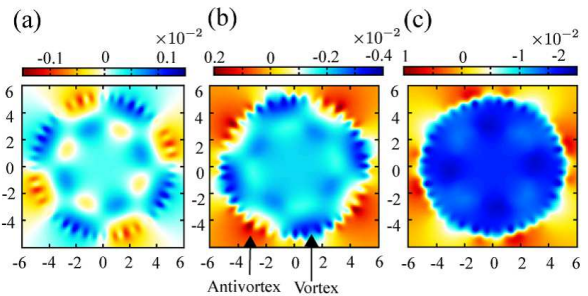}
\end{center}
\vspace{-2cm}
\caption{ESF $\tchi$ in the process after the quench.
(a) Surface ripples propagating along the surface appear first. ($t=10$.)
(b) Vortex and antivortex are generated in a pair inside the torus. ($t=11$.)
(c) Only vortices remain in the torus. SF flows are circulating in the surface
of the torus. ($t=12$.)
}
\label{chiafterQ}
\end{figure}

Finally, we see how the BEC develops towards the stable state in
Fig.~\ref{Evolution} by the stream function $\tchi(\vec{r},t)$.
First, surface instability after the sudden quench generates vortex and antivortex 
in the both sides of the boundary as show in Fig.~\ref{chiafterQ}.
Then, the calculations of $\tchi(\vec{r},t)$ show that vortices move into the BEC 
whereas antivortices leave for the outside of the BEC.
The incompressible current $\vec{j}_i$ flows counter-clock wise in the BEC
via the effect of the synthesized magnetic field.

In the following section, we shall study the instability of the dark soliton 
and formation of snake soliton by using the ESFs.


\section{Extended stream functions and snake instability of dark soliton} 

\begin{figure*}[t]
\centering
\begin{center}
\includegraphics[width=12cm]{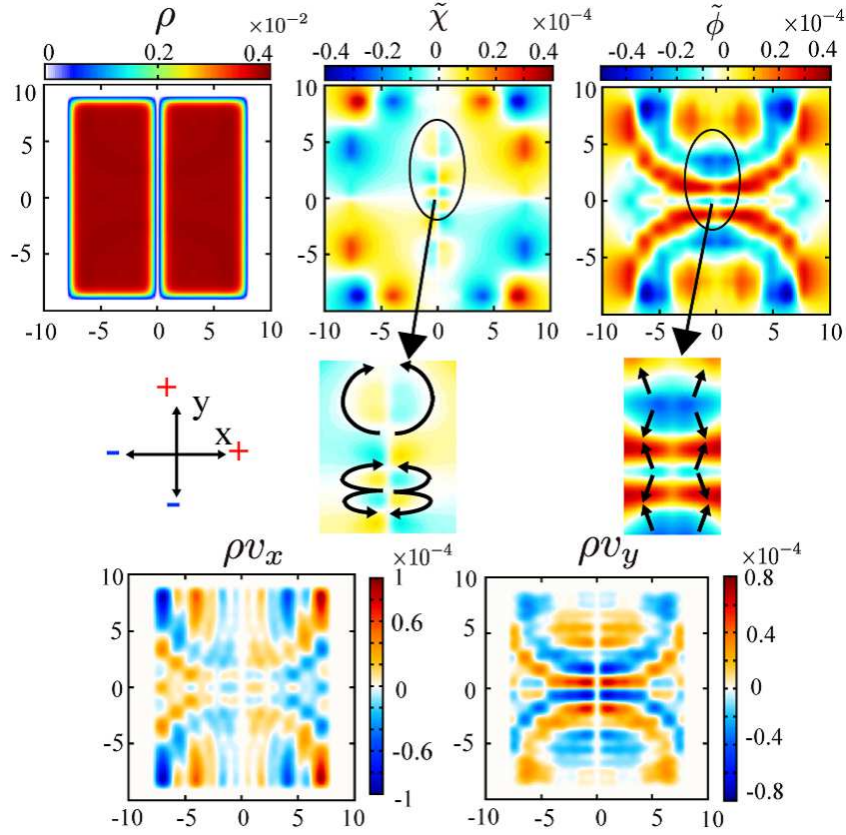}
\end{center} 
\vspace{-1cm}
\caption{BEC density $\rho$, ESFs $\tchi$ and $\tPhi$, and velocity in the first
stage of the destruction process of the dark soliton located at the center of 
the potential. ($t=3.0$.)
The arrows indicate the incompressible and compressible components of the SF velocity.
The velocity exhibits strong dependence of the $y$-coordinate, and it induces
a primordial movement of the snake instability to the second stage.
}
\label{1stst}
\end{figure*}

In this section, as an example of the dynamical behavior of BEC described by the ESFs,
we study the instability of a dark soliton located in the center of a rectangular 
confinement potential.
In GPE in Eq.~(\ref{GPeq}), we put $\omega=0$ and 
\be 
V(x,y) =\left\{
\begin{array}{ccr}
0, & (|x|<8, |y|<9) \\
100, & (\mbox{otherwise}).
\end{array}
\right .
\ee 
The initial configuration at $t=0$ is produced by imprinting the dark soliton profile
on the homogeneous BEC state such as 
$\psi_0(x)=n_0 \tanh (\mu x)$ with constants $n_0$ and $\mu$~\cite{DSWF}.
We study how this dark soliton evolves by the ESFs.

\vspace{-1cm}
\begin{figure*}[t]
\centering
\begin{center}
\includegraphics[width=12cm]{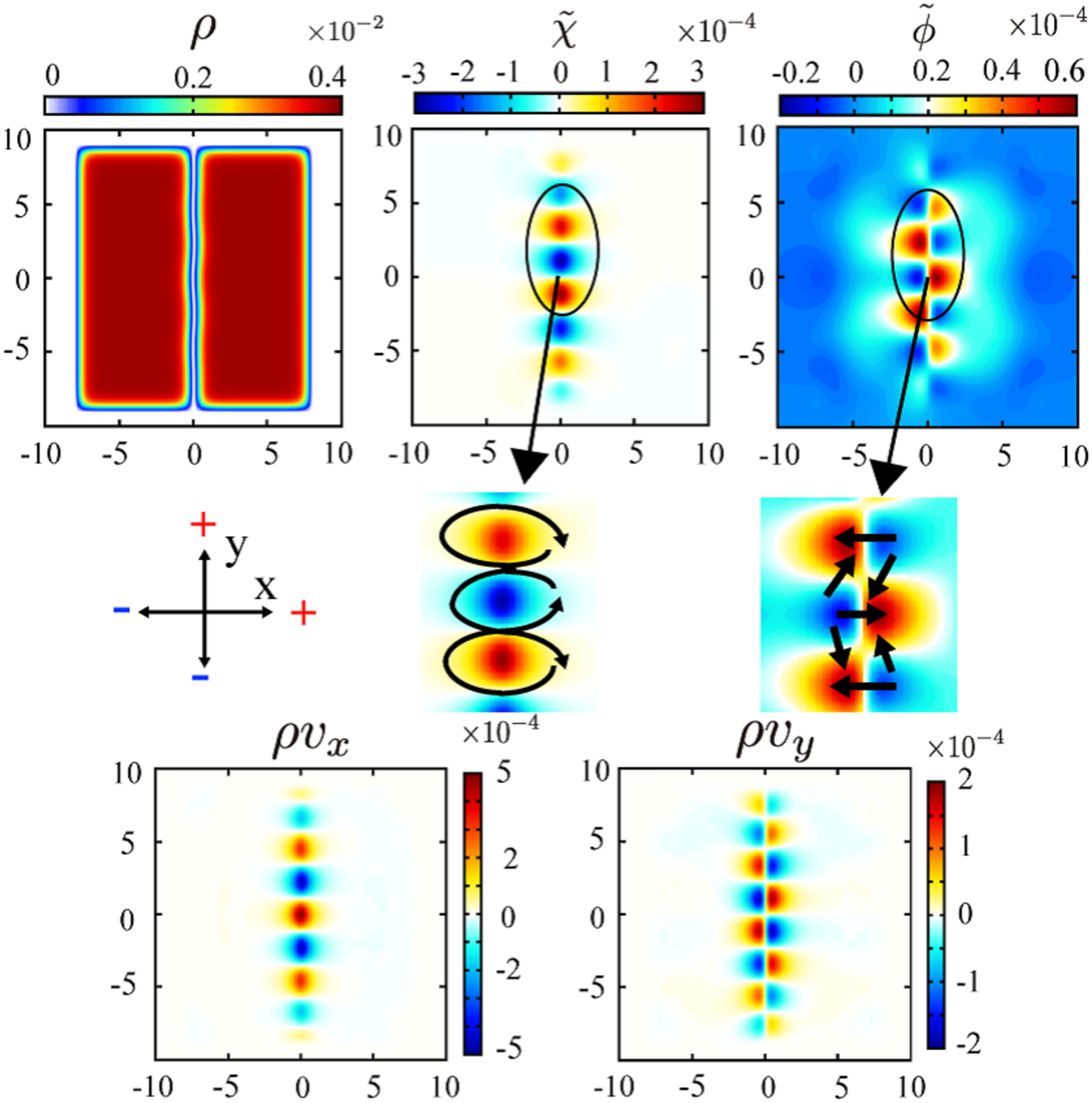}
\end{center} 
\caption{BEC density $\rho$, ESFs $\tchi$ and $\tPhi$, and velocity in the second
stage of the destruction process of the dark soliton located at the center of 
the potential. ($t=20.1$.)
The arrows indicate the incompressible and compressible components of the SF velocity.
The velocity exhibits a strong dependence of the $y$-coordinate.
}
\label{2ndst}
\end{figure*}
\begin{figure*}[t]
\begin{center}
\vspace{-2cm}
\includegraphics[width=7cm]{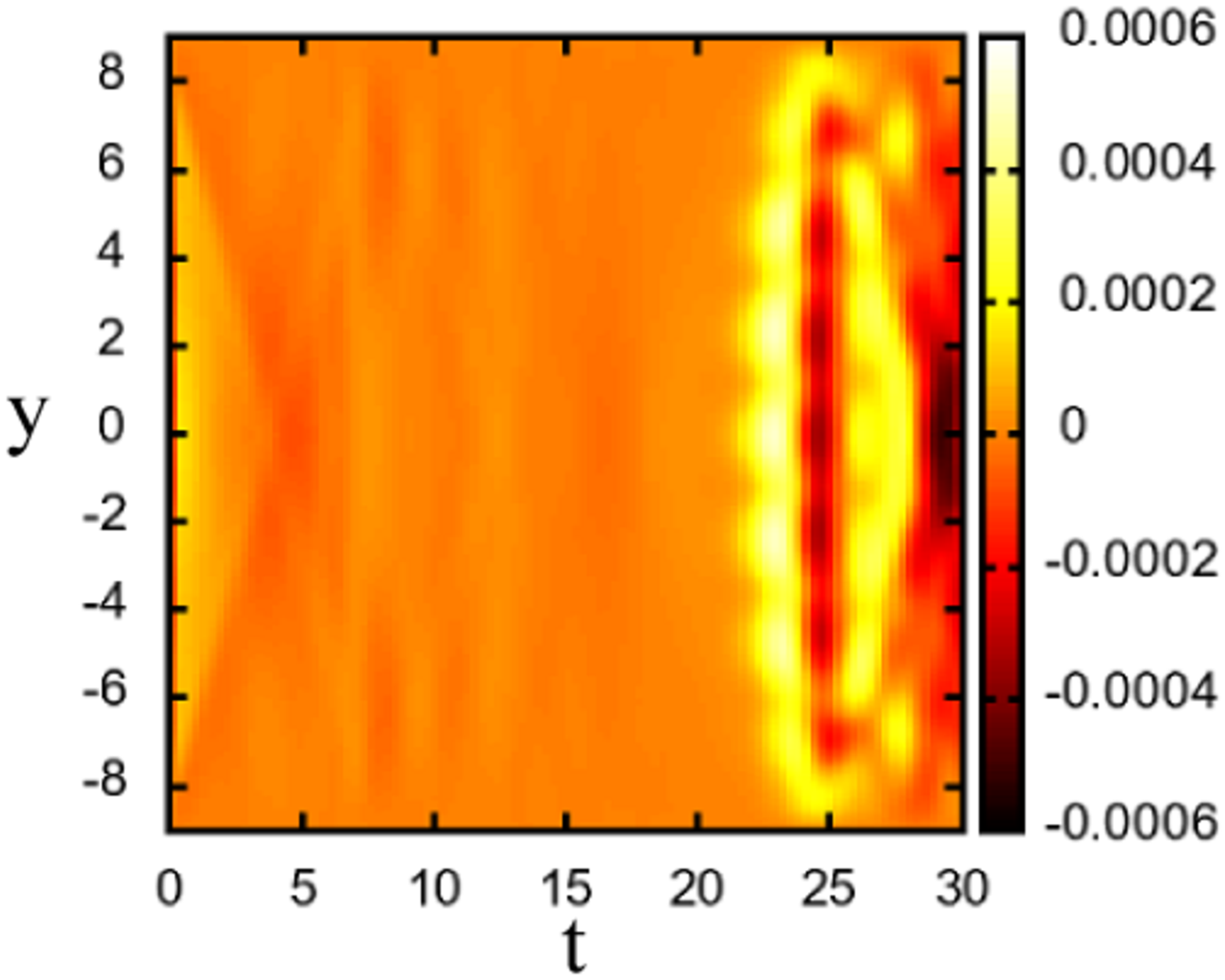} 
\includegraphics[width=7cm]{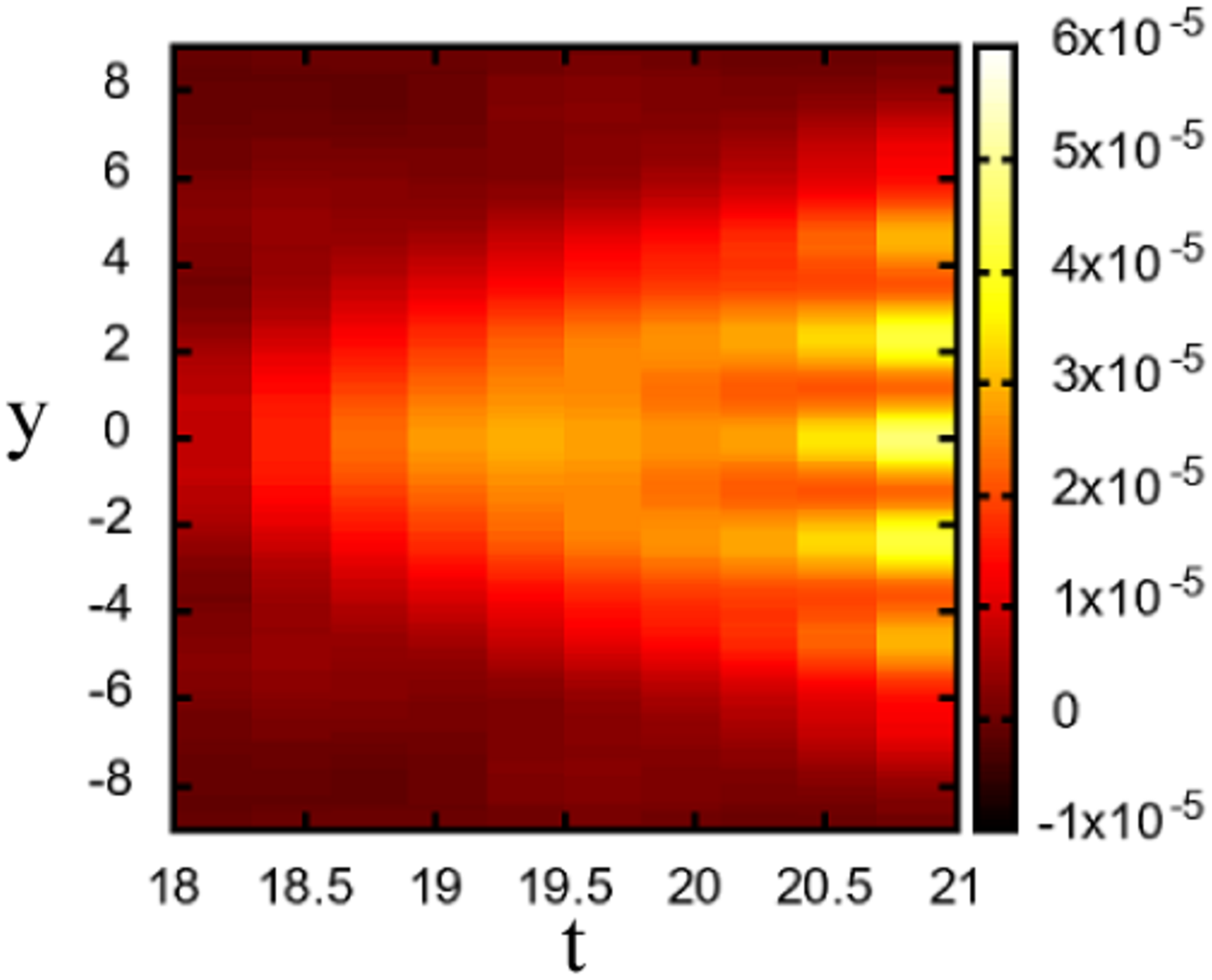} \\
\includegraphics[width=7cm]{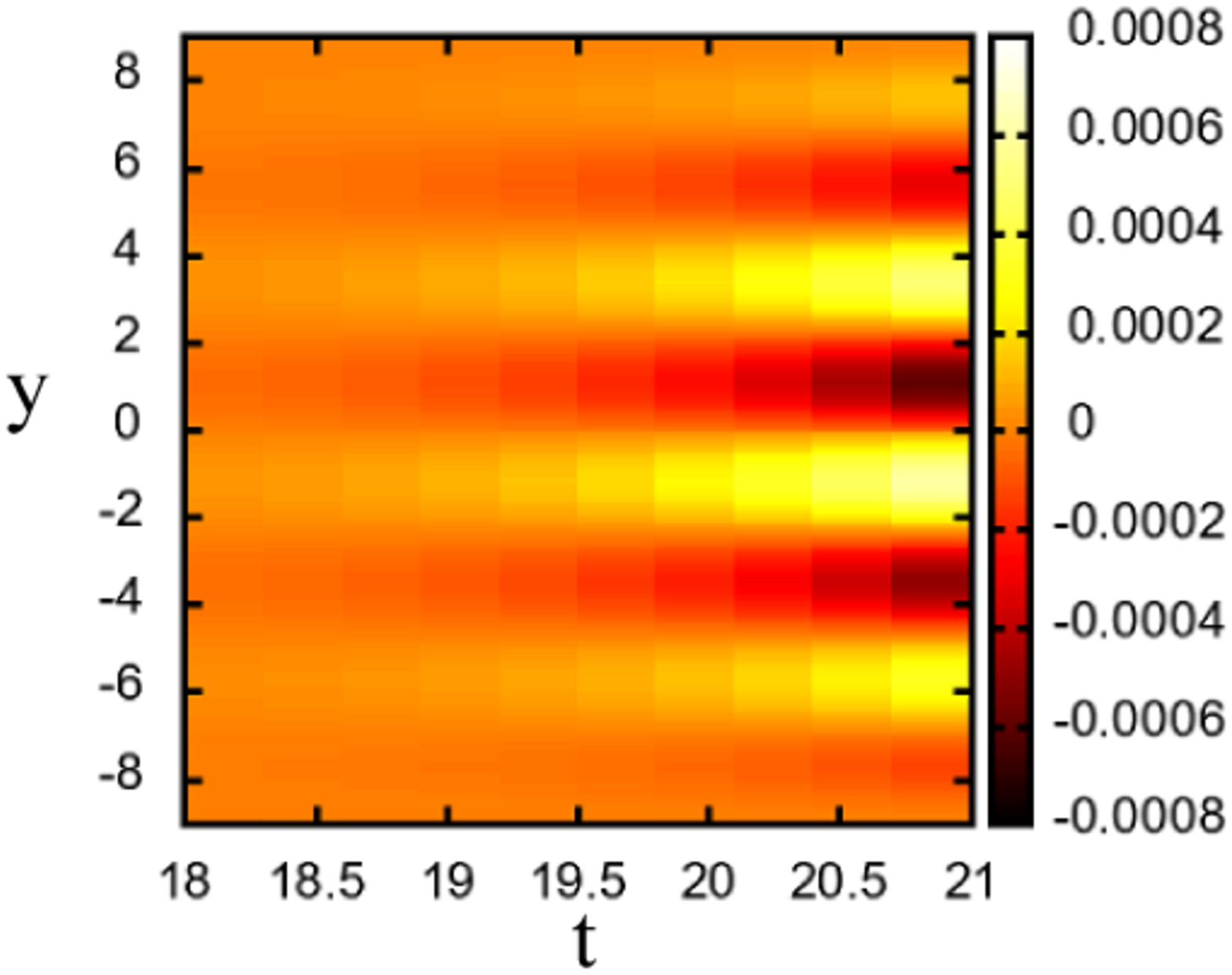} 
\includegraphics[width=7cm]{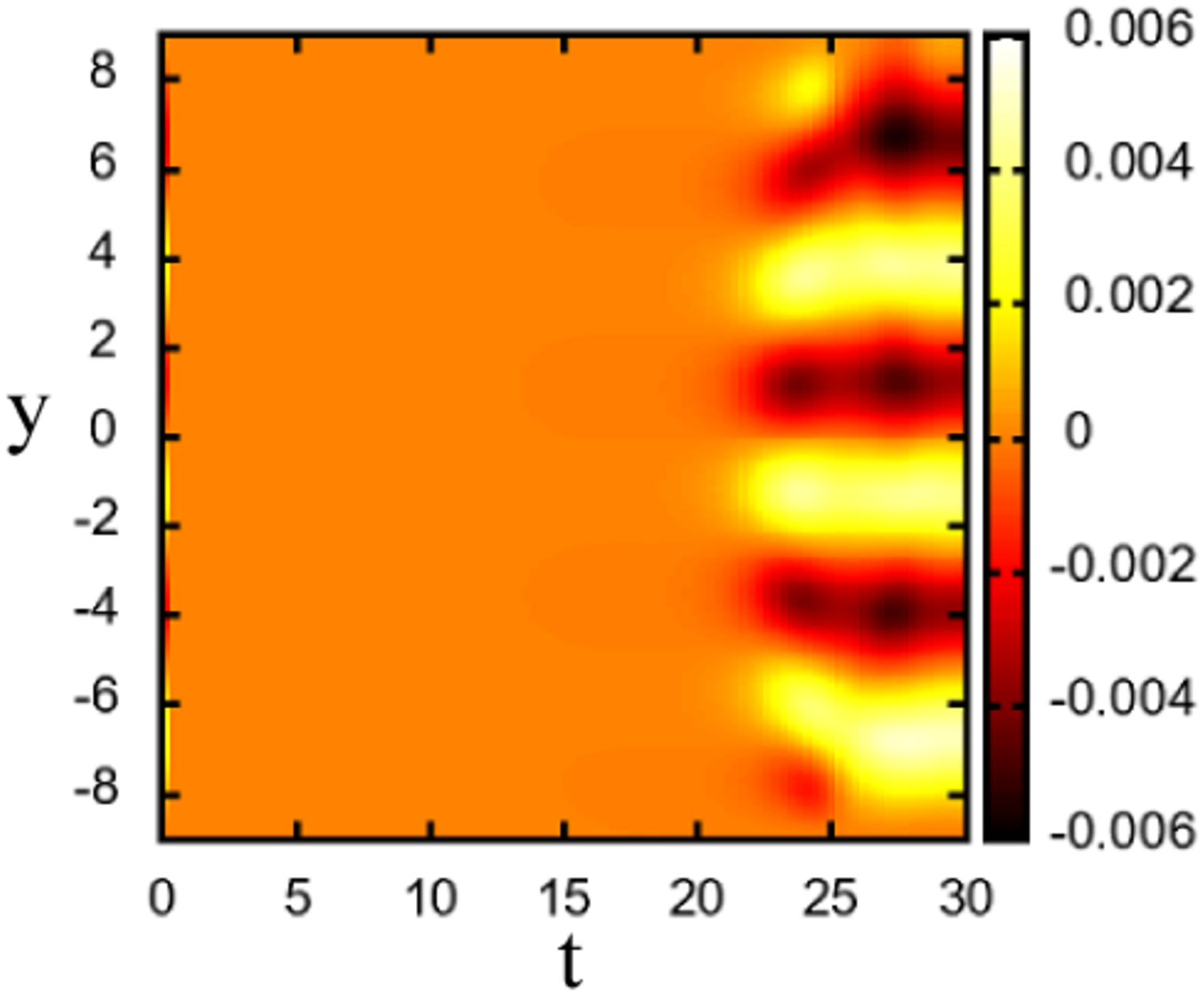}
\end{center}
\vspace{-1cm}
\caption{Upper panels: BEC flow $Q(y,t)$ in Eq.~(\ref{densityF}) that measures 
amount of SF-density flow from $y=-9$ to $y$.
Lower panels: BEC flow through the $x=0$ line, $R(y,t)$, in Eq.~(\ref{densityF2}) 
that measures amount of the incompressible SF flow from $y=-9$ to $y$.
The both results indicate the snake instability to the stage 2.
}
\label{streamline}
\end{figure*}
\begin{figure*}[b]
\centering
\begin{center}
\includegraphics[width=12cm]{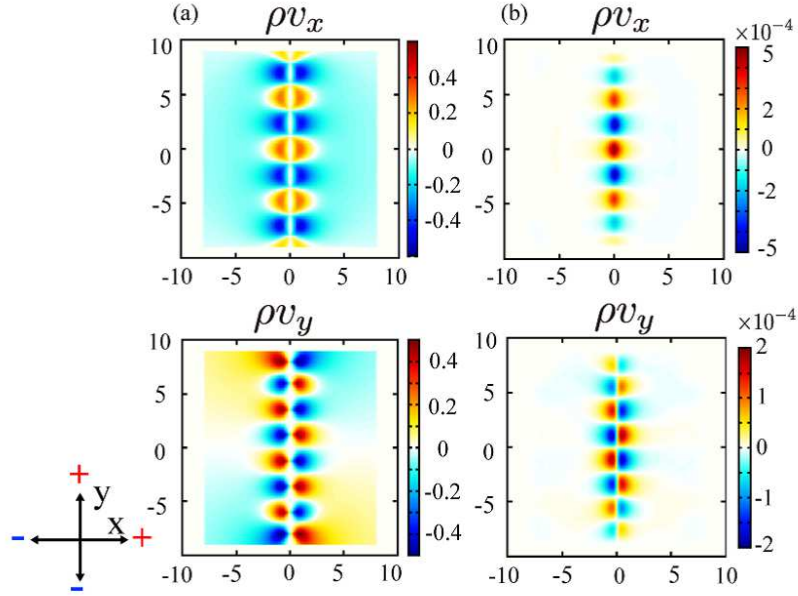}
\end{center}
\caption{Vortex-antivortex configurations.
(a) The flow profiles obtained by the analytical expression 
in Eq.~(\ref{VDflow}). (b) That obtained by the numerical simulation
in the process of the snake instability of the dark soliton.
These two are rather similar, but the configuration (b) has nontrivial
flows only in the central region. This comes from the fact that 
the configuration (b) forms in the process of the dark soliton instability.
}
\label{VDs}
\end{figure*}

\vspace{1cm}

The continuity equation is given as follows,
\be 
{\partial (\rho_0(\vec{r})+\delta \rho(\vr, t)) \over \partial t}  
&=&{\partial \delta \rho(\vr, t) \over \partial t}  \nonumber \\
&=&-\nabla\cdot \vec{j}(\vr, t) \label{cont3} \\
&=&-\nabla\cdot ((\rho_0(\vr)+\delta \rho(\vr, t))\delta \vec{v}(\vr, t)),  \nonumber 
\ee
where $\rho_0(\vr)=\psi^2_0(x)$.
On the other hand by using the ESFs,
\be
(\rho_0(\vr)+\delta \rho(\vr, t))\delta \vec{v}(\vr, t)=\nabla\tPhi
+\hat{z}\times\nabla \tchi.
\label{EFF2}
\ee
By substituting $\rho_0(\vr)=\psi^2_0(x)$ into Eq.~(\ref{EFF2}) and ignoring
term such as $\nabla\delta\rho\times \delta\vec{v}$, we obtain
\be
\hspace{-0.5cm}
\Delta\tchi(\vr,t)&=&2n_0^2\mu{\tanh(\mu x) \over \cosh^2(\mu x)}
\delta v_y(\vr,t)
\nonumber \\
&&+(\rho_0(\vr)+\delta \rho(\vr, t)) \nabla \times \delta\vec{v}.
\label{tchieq}
\ee
In the second term on the RHS of Eq.~(\ref{tchieq}), for a configuration of vortex
at $\vr=\vr_0$, $\nabla \times \delta\vec{v}\propto\delta(\vr_0)$.
However in this case, $(\rho_0(\vr_0)+\delta \rho(\vr_0, t))\simeq 0$, and therefore
this term is negligibly small.
Similarly for $\tPhi(\vr,t)$,
\be
\Delta \tPhi(\vr, t)&=&2\mu n_0^2{\tanh(\mu x) \over \cosh^2(\mu x)}\delta v_x(\vr,t)
\label{tPhieq} \\
&&+n_0^2 \tanh^2(\mu x)\nabla\cdot\delta\vec{v}(\vr,t)
+\delta \rho\nabla\cdot\delta\vec{v}(\vr,t). \nonumber
\ee 

For the first stage of the evolution with $t\ll 1$,
the numerical calculations of $\rho(\vr,t)$, $\tchi(\vr, t)$ and $\tPhi(\vr, t)$, etc.
by using Eq.~(\ref{Fourier}) are shown in Fig.~\ref{1stst}.
The BEC is approximately symmetric under $x\to -x$, 
\be 
&&\delta \rho(x,y,t)\simeq \delta \rho(-x,y,t), \nonumber \\
&&\delta v_x(x,y,t)\simeq -\delta v_x(-x,y,t), \label{initial} \\
&&\delta v_y (x,y,t)\simeq \delta v_y(-x,y,t).
\ee
In this case, $\tchi(\vr,t)$ is anti-symmetric under $x\to -x$,
as observed by the numerical calculation in Fig~\ref{1stst}.
Furthermore, the numerical calculations in Fig.~\ref{1stst} show that $\delta v_y$ 
in the region $x\sim 0$ has strong $y$ dependence as the both compressive and
incompressible components do.
On the other hand on the line $x=0$,
$\delta v_x=0$, whereas ${\partial \delta v_x \over \partial x}\neq 0$ by
the incompressible component, and therefore 
the second term on the RHS of Eq.~(\ref{tPhieq}) contributes to $\Delta\tPhi(\vr,t)$
in the small but finite $x$ region.
The relatively large $\nabla\cdot\delta \vec{v}$ induces 
a density modulation $\delta \rho$ through Eq.~(\ref{tPhieq}) 
and the continuity equation Eq.~(\ref{continue2}).
This is a primordial movement of the snake instability. 

\begin{figure*}[h]
\centering
\begin{center}
\includegraphics[width=12cm]{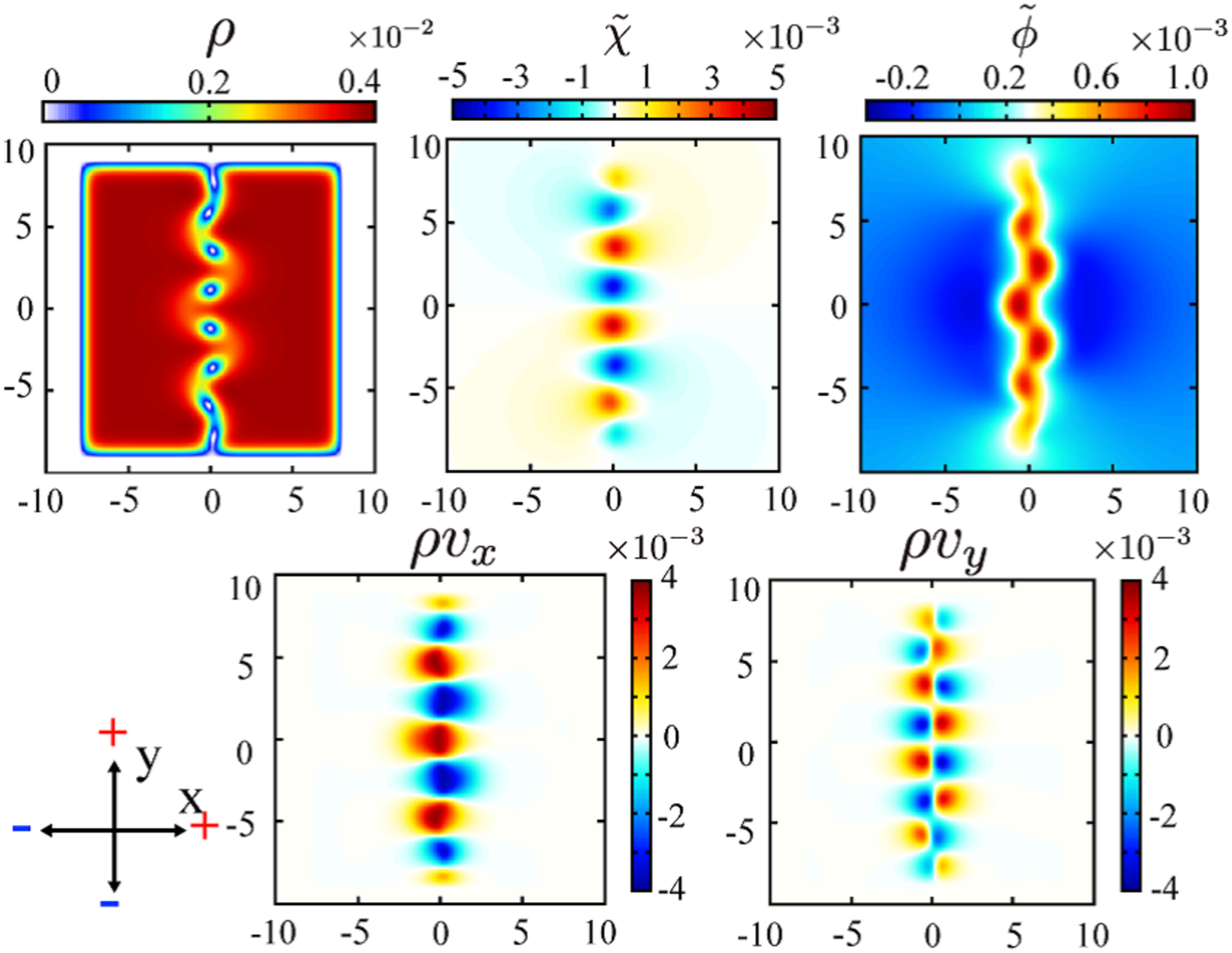}
\end{center}
\caption{BEC density $\rho$, ESFs $\tchi$ and $\tPhi$, and velocity in the third
stage of the destruction process of the dark soliton located at the center of 
the potential. ($t=23.4$.)
The velocity exhibits a strong dependence of the $y$-coordinate.
}
\label{3rdst}
\end{figure*}
\begin{figure*}[h]
\centering
\begin{center}
\includegraphics[width=12cm]{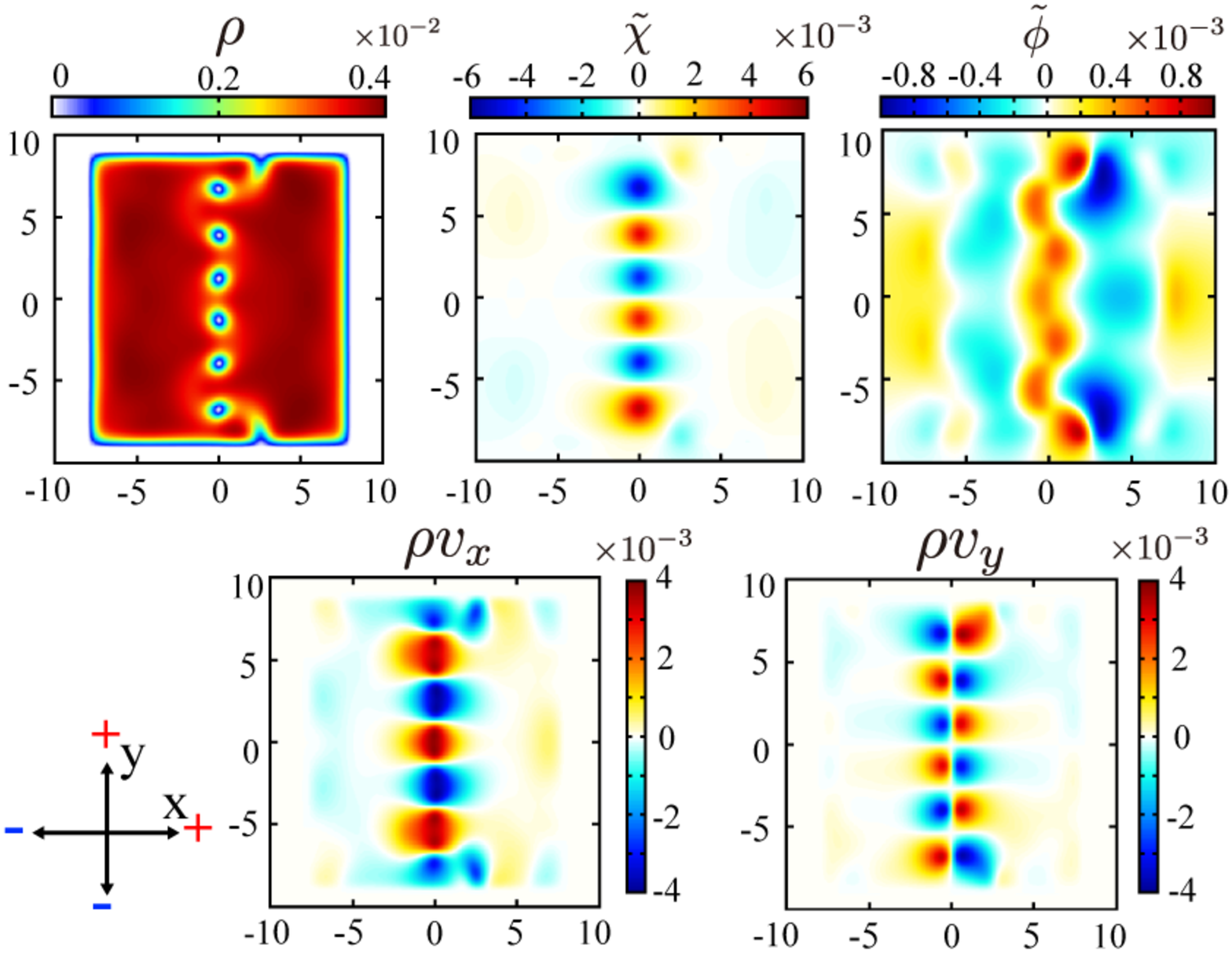}
\end{center}
\caption{BEC density $\rho$, ESFs $\tchi$ and $\tPhi$, and velocity in the fourth
stage of the destruction process of the dark soliton located at the center of 
the potential. ($t=27.0$.)
Isolated vortex droplets form and this fourth stage is stable against the pair 
annihilation of vortex and antivortex.
}
\label{4thst}
\end{figure*}

In the second stage of the evolution, the numerical results of the ESFs 
in Fig.~\ref{2ndst} reveal that $\delta v_x$ comes from both the compressible and
incompressible flows, whereas $\delta v_y$ from the compressible component.
As $\tPhi$ in Fig.~\ref{2ndst} shows, the central empty line slightly deforms 
to a meandering line by the density flow explained above.

By the numerical simulation, we observed that the second stage is 
a meta-stable state.
It is interesting to see how it forms from the BEC in the first stage.
In Fig.~\ref{streamline}, we show 
the incompressible SF density flow at $t$, $Q(y,t)$, defined by
\be
Q(y,t)&=&\int^y_{y_0} \rho(0,y',t)\vec{v}(0,y',t)\cdot \hat{y} dy' \nonumber \\ 
&=&\tPhi(0,y,t)-\tPhi(0,y_0,t),
\label{densityF}
\ee
where $\hat{y}$ in the unit vector in the $y$-direction, and we put $y_0=-9$.
From the calculations in Fig.~\ref{streamline}, it is seen that a density modulation 
starts to form at the center of the dark soliton ($y\simeq 0$) at $t\sim 18$ 
and then it spreads in the $y$-direction.
Similarly, we show the SF incompressible flow in Fig.~\ref{streamline}, $R(y,t)$, which is defined as 
\be
R(y,t)&=&\int^y_{y_0} \rho(0,y',t)\vec{v}(0,y',t)\cdot \hat{x} dy' \nonumber \\ 
&=&\tchi(0,y,t)-\tchi(0,y_0,t).
\label{densityF2}
\ee
$R(y,t)$ shows that the snake instability is enhanced by the incompressible SF 
flow as well as the incompressible flow.

It is also interesting to see how the vortices and antivortices form
in the central-meandering low-density region in the second stage.
To observation this, we calculate the SF current generated by vortices and
antivortices by employing analytic functions as a reference.
For example, the flow in a typical vortex-antivortex configuration 
is given as,
\be
&&\rho v_x=\tanh^2(\mu x) \Big(\sum_{i=1}^8
{-s_i(y-y_i) \over x^2+(y-y_i)^2}\Big),
\nonumber \\
&&\rho v_y=\tanh^2(\mu x) \Big(\sum_{i=1}^8
{s_i(x-x_i) \over x^2+(y-y_i)^2}\Big)
\label{VDflow}
\ee
where the vortices and antivortices are located alternatively at 
$(x_i,y_i)$ with $x_i=0$ and $y_i=8,6,3.6,1.25,-1.25,-3.6,-6$ and $-8$,
and the vorticity $s_i$ is given by $s_i=(-)^i$.
Configuration of the above flow (\ref{VDflow}) is shown in Fig.~\ref{VDs}, 
which exhibits a qualitative similar profile to that in Fig.~\ref{2ndst} in the central region, 
but the configuration given by Eq.~(\ref{VDflow}) has a nontrivial 
flows in the wider region compared to the results in Fig.~\ref{2ndst}.
This difference comes from the fact that vortex configuration in Fig.~\ref{2ndst} 
have just formed in the second stage, and therefore the region far apart from 
the central region is not influenced by the vortex formation.

The density flow in Fig.~\ref{2ndst} continues after the second stage, and then the empty
line in the central region of the SF splits into smaller lines and droplets.
This is the third stage in Fig.~\ref{3rdst}.
This behavior continues to the fourth stage in Fig.~\ref{4thst}, in which a line of droplets
form.
Each droplet corresponds to vortex or antivortex as the ESF, 
$\tchi(\vr,t)$, indicates. 
That is, vortices and antivortices align in a straight line alternatively.
We observed that this fourth stage is rather long-lived, even though
a pair annihilation of vortex and antivortex is possible.


\section{Conclusion}

In this work, we introduced the ESFs to describe dynamics of non-stationary and
compressible SF.
To this end, two components of the ESFs, $\tchi$ and $\tPhi$, are necessary.
By the definition of the ESFs, vortices can be described with regular $\tchi$ and $\tPhi$,
and also specific boundary conditions are not necessary in the confinement
potential problems.
We showed the numerical methods to obtain the ESFs, and applied the ESFs to 
the sudden quench 
of SF in the torus by applying a synthetic magnetic field.
We found good agreement between the results obtained by our methods and 
those by the direct calculations from the solutions to the GP equation.

Then, we applied our methods to detailed study on the snake instability of 
the dark soliton in the  rectangular potential.
We found that there exist four typical stages in the process, and clarified dynamics
of the SF flow in the whole process of the instability towards the meta-stable state
with vortex droplets.

We hope that the ESFs given in the present work are useful for study on 
various dynamics of SF.
This problem is under study, and we hope that results will be reported in the near future.

\clearpage


\end{document}